# Climate Immobility Traps:
# A Household-Level Test[*]

Marco Letta[§], Pierluigi Montalbano[†], Adriana Paolantonio[‡]

## Abstract

*The complex relationship between climate shocks, migration, and adaptation hampers a rigorous understanding of the heterogeneous mobility outcomes of farm households exposed to climate risk. To unpack this heterogeneity, the analysis combines longitudinal multi-topic household survey data from Nigeria with a causal machine learning approach, tailored to a conceptual framework bridging economic migration theory and the poverty traps literature. The results show that pre-shock asset levels, in situ adaptive capacity, and cumulative shock exposure drive not just the magnitude but also the sign of the impact of agriculture-relevant weather anomalies on the mobility outcomes of farming households. While local adaptation acts as a substitute for migration, the roles played by wealth constraints and repeated shock exposure suggest the presence of climate-induced immobility traps.*

**JEL-Codes:** C31; O15; Q54.

**Replication package:** Available at the [World's Bank Reproducible Research Repository](#).

**Declarations of interest:** None.

[§] Department of Social Sciences and Economics, Sapienza University of Rome; marco.letta@uniroma1.it
[†] Department of Social Sciences and Economics, Sapienza University of Rome; pierluigi.montalbano@uniroma1.it
[‡] Development Data Group, World Bank; apaolantonio@wordbank.org

[*]For valuable comments, we are grateful to Gero Carletto, Augusto Cerqua, Talip Kilic, Kevin McGee, Donato Romano, Asmelash Haile Tsegay, Alberto Zezza and seminar participants at the World Bank. We also thank Thomas Patrick Bentze, Pauline Castaing, Josephine Durazo, Siobhan Murray, Federica Petruccelli, Akiko Sagesaka, and Philip Randolph Wollburg.

*"Our understanding of the links of 1.5°C and 2°C of global warming to human migration is limited and represents an important knowledge gap"*

Special Report of the Intergovernmental Panel on Climate Change (Hoegh-Guldberg et al., 2018)

1. Introduction

The field of climate migration faces a paradox: climate mobility is increasingly under the media and political spotlights, yet scientific evidence on this phenomenon is mixed and inconclusive, resulting in a lack of clear-cut policy prescriptions. For instance: who are the climate migrants? To date, this apparently simple question lacks a definitive answer.[1] An even deeper layer of uncertainty envelops the mediating role played by rural development and adaptation policies: will they increase climate-induced mobility—by relaxing financial liquidity constraints that prevent people from leaving—or decrease it by enabling on-farm adaptation, thus making migration unnecessary? Answering such questions requires a deep understanding of farmers' mobility dynamics under climate risk and of the transmission mechanisms that allow identification and targeting of different groups. While a better understanding of the dynamic relationship between local adaptation and migration is a central issue, there has been little research on the matter, resulting in a need for joint modeling of different mobility and adaptation outcomes (Cattaneo et al., 2019). In this work, we propose new ways to uncover the relationship between climate shocks, (im)mobility, and adaptation, and, in turn, inform policymaking.

In a recent review, Letta et al. (2023) find that household responses to climate-related shocks are irreducibly heterogeneous, context-dependent, and driven by a host of socio-economic channels, to the extent that, depending on the role played by these intervening factors, a weather anomaly can equally lead either to an increase or to a decrease in household mobility. This implies that focusing on average effects is misleading when assessing climate-induced migration, since averages wash away the substantially heterogeneous responses of structurally different groups. Without thoroughly unpacking this heterogeneity puzzle, it is implausible to make statements *a priori* on current patterns of climate mobility, let alone future projections. It follows that rigorously estimating heterogeneous treatment effects is an essential prerequisite for the optimal targeting of climate and development policies.

---

[1] This is particularly relevant for policymaking as the concept of 'climate migrant' breaks the standard dichotomy between economic migrants and refugees from both a normative and a legal point of view. In this regard, the World Bank's *World Development Report 2023* acknowledges that there is no clear-cut distinction between economic migrants and refugees, but rather a continuum of complex patterns of movements (or lack thereof) corresponding to varying degrees of protection needs (World Bank, 2023).

Guided by these reflections, we couple a causal machine learning approach based on a data-driven search for treatment effect heterogeneity (Athey & Imbens, 2016; Wager & Athey, 2018), tailored to a conceptual framework bridging the New Economics of Labor Migration (NELM) and the poverty traps literature, with longitudinal, multi-topic household survey data collected in Nigeria with support from World Bank's Living Standard Measurement Study (LSMS). When we refer to a 'data-driven search for treatment effect heterogeneity', we mean that we go beyond estimating average treatment effects. Instead, we calculate the impacts of weather shocks at the household level. These individual impacts are then aggregated into group-average treatment effects. Importantly, the groups are identified through data-driven sample splits based on specific household characteristics before the shock. Our non-parametric estimates of treatment effect heterogeneity suggest that some key variables —namely, pre-shock asset levels, *in situ* adaptive capacity, and cumulative exposure to weather shocks—can alter not just the magnitude, but even the sign of the causal impact of growing season weather anomalies on the (im)mobility outcomes of Nigerian agricultural households. Overall, our analysis provides empirical evidence that suggests the presence of climate-induced immobility traps.

Among recent works related to this paper, Martinez Flores, Milusheva and Reichert (2021), leverage high-frequency data collected from the International Organization for Migration in seventeen West and Central African countries over the period 2018-2019, and estimate that droughts occurring during the growing season reduce international migration. As they find these effects only in middle-income areas (and not in rich nor poor ones), they infer that liquidity constraints and income losses are the key channels. Mueller et al. (2020) use data from LSMS-supported longitudinal surveys conducted in Ethiopia, Malawi, Tanzania, and Uganda over the period of 2009-2014 and find that climate variability and weather anomalies cause a reduction in temporary urban migration in Eastern Africa. In contrast, Di Falco et al. (2023) also use data from LSMS-supported longitudinal surveys conducted in Ethiopia, Malawi, Niger, Nigeria, and Uganda to analyze the effects of cumulative drought shocks on the migration decisions of rural households and find that a positive effect of multiple droughts accumulates over time. A study by Dillon et al. (2011) uses data from Northern Nigeria finds that households respond to agricultural risk by sending males to migrate. None of these studies, however, carry out a systematic search for the drivers underlying the heterogeneity of climate (im)mobility outcomes, nor tries to disentangle the relationship between migration and local adaptation. Furthermore, a methodological limitation common to all these works is that they employ simple linear probability models, which not only impose restrictive functional forms, but by design can only estimate average treatment effects and check whether they differ across pre-specified subgroups; they do not allow for arbitrary and unrestricted treatment effect heterogeneity nor for nonlinear responses.

We provide three contributions to the existing research. The first is substantive: we produce data-driven evidence on the key channels that mediate the microeconomic relationship between climate shocks and mobility, providing empirical support for the hypothesis of climate immobility traps. This is a particularly relevant notion from a policy perspective as it implies the existence of climate-induced wealth and liquidity constraints. The evidence we document is in line with recent meso- and macro-evidence on the topic (Benveniste et al., 2022; Cattaneo & Peri, 2016; Peri & Sasahara, 2019) as well as with more general findings that the most vulnerable are often not those who migrate, as they are too credit-constrained and lack the necessary resources to do so (Cai, 2020; Dustmann & Okatenko, 2014). An increasing number of studies suggests that policymakers should be concerned about immobility as much as mobility (Findlay, 2012; Cattaneo et al., 2019). Indeed, it is now commonly acknowledged that too often the policy focus has been placed on 'those who leave' rather than on 'those who cannot leave', despite the evidence that climate change is unlikely to trigger mass migration, at least in the near future (Boas et al., 2019; Burzyński et al., 2022). Our paper goes a step further by shedding light on the key drivers of the so-called "immobility paradox" (Beine et al., 2021), i.e., that fewer people migrate due to climate change than is otherwise expected. Overall, the empirical evidence we provide makes a strong case for targeted measures and group-specific, rather than 'catch-all', agricultural programs and adaptation policies.

The second contribution is methodological: we are not aware of other studies applying causal machine learning to open the black box of climate-induced migration. Specifically, we propose a systematic, rather than *ad hoc*, approach to shed light on the irreducible heterogeneity of climate-induced mobility outcomes. Our approach improves upon traditional *ad hoc* subsample analysis (or use of interaction terms) typically employed in the specialized literature to uncover underlying heterogeneity because it limits researchers' discretion when selecting the relevant dimensions and thresholds of heterogeneity. Given such heterogeneity, traditional econometric tools for the estimation of average treatment effects are not fit for purpose, at least not without strong *a priori* assumptions by the analyst. The tree-based machine learning techniques we employ (Athey & Imbens, 2016; Wager & Athey, 2018) are, instead, agnostic regarding functional forms, easily capture nonlinearities and interactions among key household characteristics, and identify the main heterogeneity drivers and thresholds in a data-driven manner. This enables us to estimate policy relevant group-average treatment effects and show how average impacts can be misleading, since they conceal substantial heterogeneity, with opposite effects for different groups. Importantly, given the multidimensionality of the migration phenomenon, this approach presupposes that the researcher has access to longitudinal multi-topic and multi-purpose household survey data, which collect information about the most important mediation channels and transmission mechanisms, including agricultural variables.

The last contribution is conceptual: we frame and explore the climate-migration nexus through the lens of the poverty traps approach. Specifically, we provide an extension of the asset-based and geographic poverty traps literature (Carter & Barrett, 2006; Barret & Carter, 2013; Jalan & Ravallion, 2002) to the immobility framework, coupled with a perspective of migration phenomenon as a household-level risk management strategy inspired by the NELM (Stark & Bloom, 1985). Such a framework serves as the domain knowledge basis for our empirical approach aimed at detecting potential regime shifts due to the existence of "climate immobility thresholds", below which households fall into immobility traps. If trapped populations are concentrated in areas plagued by recurrent shocks, this may, in turn, give rise to geographic poverty traps, which have been assessed as the most likely form of poverty trap by leading scholars and one for which policies promoting migration might prove particularly beneficial (Kraay & McKenzie, 2014).

Lastly, in addition to the aforementioned, our analysis holds implications for enhancing the design of future longitudinal surveys to produce more and better data for addressing critical gaps in climate migration research. Desired improvements concern the collection of relevant contextual information regarding the 'migrants' in order to accurately classify different types of migration and unravel more clearly their drivers and impacts. This includes details about the nature and duration of migration episodes, as well as the characteristics, history, and purposes of migrant individuals. By gathering such additional information and carefully tracking migrant individuals over time, we could create migration variables that go beyond merely measuring the presence or absence of migrants within a household and provide more nuanced evidence to better inform targeted policies and programs.

The rest of the paper is arranged as follows. Section 2 outlines the conceptual framework. Section 3 presents data sources and descriptive statistics, while Section 4 describes the methodology. Section 5 illustrates the results of the empirical analysis and Section 6 concludes.

## 2. Conceptual framework

Our conceptual framework is placed at the crossroads of the NELM and poverty traps literatures. The distinctive features of the NELM theory, in contrast to the neoclassical counterfactual, lie in its consideration of migration decisions as household-driven rather than individual-oriented in the presence of incompleteness in insurance and credit markets (Stark & Bloom, 1985). Moreover, the objective function is not limited to maximizing expected income, but also includes minimizing risk exposure (Millock, 2015). Within this framework, where formal risk management tools are absent, households manage risk by strategically allocating family workers to diverse and uncorrelated (i.e., geographically dispersed) labor markets with the aim of diversifying income streams (e.g., via remittances). In the context of rational behavior models, households facing higher levels of risk assess the expected remittances against the marginal contribution of each family member to local agricultural productivity under stress. This establishes a connection between migration and agricultural productivity, the direction of which is context-specific and depends—in addition to other determinants such as factor endowments, technology, etc.—on the relative scarcity or abundance of agricultural workers. The poverty traps literature, on the other hand, delves into the self-reinforcing mechanisms whereby households risk being pushed below the poverty trap threshold after a shock (Barrett and Carter, 2013; Barret et al., 2019). The seminal work by Carter and Barrett (2006) elucidates the conditional expectation function governing the welfare dynamics of households in the aftermath of a shock, conditioned on their initial endowments and liquidity constraints.

Based on the above considerations, we elaborate a series of hypotheses regarding differential scenarios of climate-induced mobility and welfare trajectories of farming households living in rural developing contexts. Figure 1, inspired by a key study in the poverty traps literature (Carter et al., 2007), provides a stylized representation of households' welfare trajectories in the face of a series of weather shocks (lower panel) that can erode their endowment base, composed by assets and overall adaptive capacity (upper panel).

of adaptation options, which include local adaptation strategies (e.g., using irrigation, switching to drought-tolerant seeds, diversification of livelihoods away from agriculture, buying an index-based insurance contract) or sending one or more household members away to reduce agricultural income risk (*ex ante*) or losses (*ex post*).

A crucial aspect in this context is the potential existence of "immobility traps" thresholds. These thresholds, whose identification is ultimately an empirical question, are akin to the "Micawber" poverty traps thresholds proposed by Carter and Barrett (2006). Similar to the standard poverty traps dynamics, where consumption smoothing under credit rationing results in a depletion of asset levels and reduced returns, households situated below the immobility traps threshold (such as group $A_4$ in Figure 1) are prevented—due to their extremely low endowments in terms of wealth, assets, and adaptive capacity potential—from adopting mobility of household members as a risk-diversification or risk-coping option (Kraay and McKenzie, 2014, Burzyński et al., 2022).[2] Notably, even households with initially sufficient endowment levels (group $A_3$), if subject to cumulative climatic stress and repeated shocks in highly exposed rural and geographical contexts, may eventually fall into immobility and their ability to return to their convergent pre-shock

---

[2] Since risk diversification strategies such as mobility are key contributors, jointly with assets and financial liquidity, to the stability and preservation of household welfare under risk, this scenario aligns with the standard literature on poverty traps.

trajectory will be impeded.[3]

## Figure 1: Climate immobility thresholds

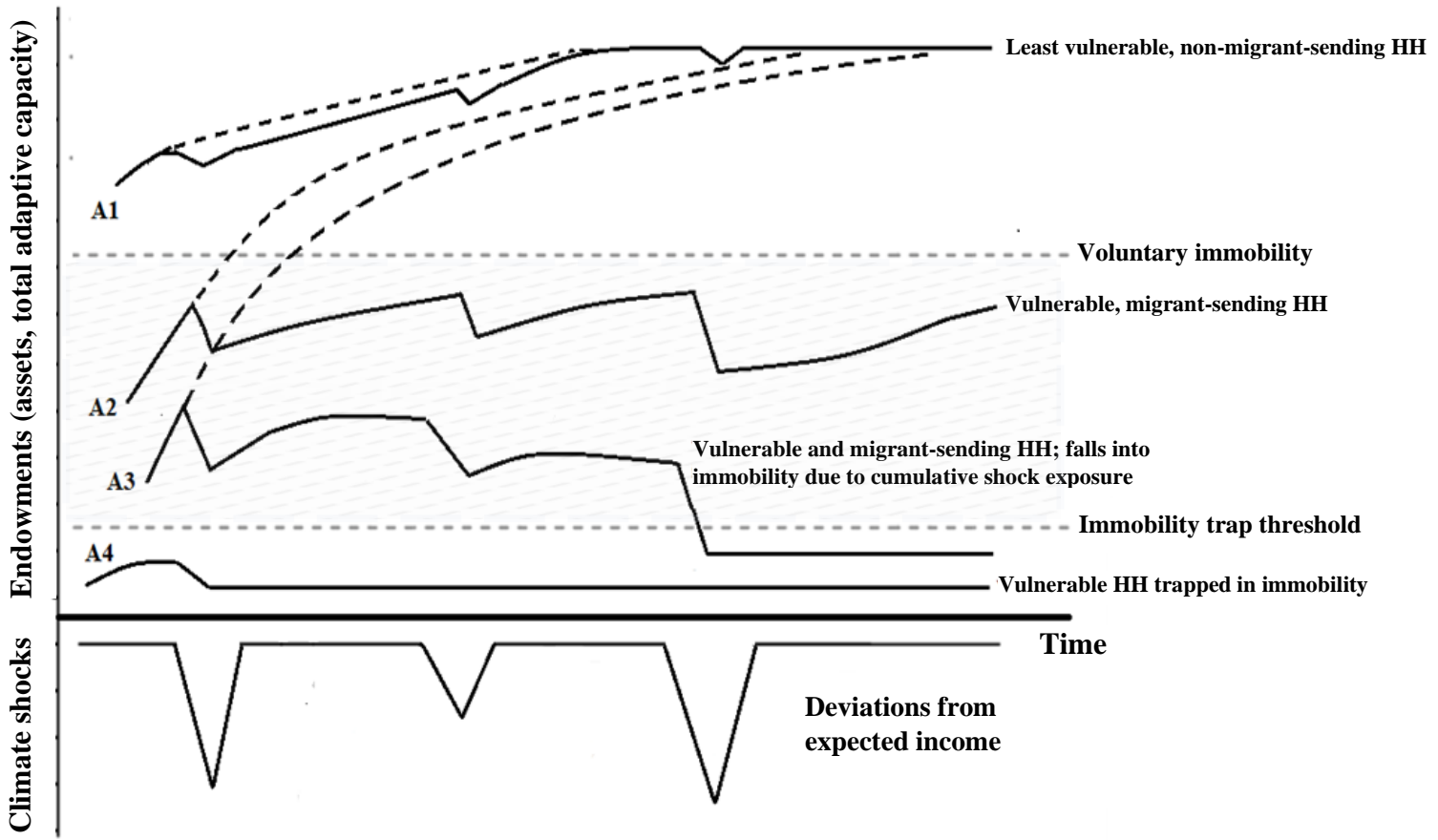

*Note*: The dashed lines represent the trajectories of the different groups in the absence of weather shocks. The shaded area between the two thresholds indicates the mobility area.

As illustrated in the figure, recurrent climate shocks are expected to hamper households' convergence over time. For example, they may slow down the convergence of the poorer

---

[3] In geographical contexts prone to recurrent and/or prolonged climate stress, weather shocks may disrupt the mechanisms through which assets (e.g., land) and other adaptation practices (including mobility, as suggested by the NELM theory) influence the expected livelihood function in a lasting manner. A pertinent example of this phenomenon can be observed in the case of the prolonged Ethiopian drought of 1998–2000 (Carter et al. (2007).

household group $A_2$ towards wealthier $A_1$ households. Over time (moving from left to right in the illustration), this is associated with heterogenous dynamics in households' welfare, conditional on key factors such as the cumulative exposure to climate shocks, initial asset and liquidity levels, and adaptive capacity.[4] To deal with climate risk, vulnerable households choose among a variety

Our argument, which will be subject to empirical scrutiny, posits that standard poverty traps thresholds can be framed as immobility thresholds when analyzed in the context of climate mobility. Climate-induced immobility thresholds are determined by a combination of households' heterogenous asset endowments, adaptive capacity potential—including local mobility patterns and off-farm employment of family members—and climate risk characteristics of the area in which they reside (i.e., the intensity and frequency of weather events determine levels of risk exposure). In this perspective, asset levels, adaptation options and choices, and cumulative shock exposure lead to bifurcation among households, whereby some of them are unable to send family members out as an adaptive mechanism due to a variety of reasons including climate-induced wealth and liquidity constraints, lack of information, and limited adaptive capacity potential and network facilities. This alteration in the convergence process, driven by the heterogeneity in household-level migration determinants depicted by the NELM approach, can result in certain groups of households remaining trapped in immobility—as illustrated by the divergent trajectories of groups $A_2$ and $A_3$ in Figure 1. Since vulnerable households below this immobility threshold tend to become trapped in high-risk locations, immobility thresholds can also lead to the emergence of climate-related poverty pockets and geographic poverty traps (Jalan & Ravallion, 2002; Kraay & McKenzie, 2014).

In addition to this immobility trap threshold, Figure 1 also depicts an upper threshold for a different type of immobility—which we label 'Voluntary immobility'—above which, for better-endowed households, mobility is no longer a needed or desirable risk management or risk coping strategy, as these richer households are barely vulnerable to climate stress and quickly recover from shocks

---

[4] In a certainty equivalent scenario, we would observe a process of convergence: diminishing returns to wealth imply that initially poorer households catch up to the welfare levels of their better-off neighbours (Carter and Barrett, 2006). This is analogous to the "conditional convergence" issue that has figured prominently in the macroeconomic debate.

without long-lasting welfare consequences or the need to resort to migration.

This conceptual framework offers valuable insights and constitutes a practical tool for analyzing climate-induced mobility. First, it enables the assessment of the inherently non-linear relationship between climate shocks and mobility. Second, it allows for a specific investigation of the key mediating factors influencing this relationship, which can be attributed to household-specific endowments and contextual elements, as in the NELM framework. Overall, it helps in identifying the heterogeneous effects of climate shocks, where outcomes include not only dichotomic choices but also encompass mobility, immobility, and poverty traps. Lastly, it emphasizes the importance of estimating critical thresholds that determine households' migration responses to climate stress based on their assets, adaptive capacity, and liquidity constraints.

It is important to note that most conventional resilience and vulnerability indicators and traditional econometric tools are not adequately equipped to capture the intrinsically non-linear dynamics of the heterogeneous (im)mobility responses induced by climate shocks depicted in Figure 1 (Montalbano and Romano, 2022). Therefore, novel empirical approaches based on fully flexible and nonparametric methods are needed to estimate the conditional probability of farmers' mobility under climate risk (as explored by Cissé and Barrett, 2018). Our empirical analysis aims to explore these issues using household panel data and methodologies capable of detecting the type of conditional treatment effect heterogeneity that, according to our framework, characterizes the climate-mobility nexus in rural agricultural contexts.

In conclusion, based on the above discussion, we formulate the key hypotheses subject to subsequent empirical testing, presented in Table 1. The table also reports the expected empirical results that one would anticipate *ex ante* based on the relationships described in the conceptual framework.

**Table 1: Theory-based hypotheses and expected findings**

|    | Hypothesis | Expected finding |
|----|------------|------------------|
| H1 | *There is a critical wealth threshold below which households fall into climate immobility traps.* | Negative impacts of weather anomalies on migration for households with lower endowments (e.g., asset levels). |
| H2 | *Households exposed to recurrent shocks are more likely to experience resource-constrained immobility.* | Concave relationship between the estimated effect of contemporaneous weather shocks on migration and lagged weather conditions. |
| H3 | *Local adaptation and migration are substitute risk management strategies that households adopt to deal with climate risk.* | Negative relationship between the estimated impact of weather shocks on migration and pre-shock levels of local adaptive capacity. |
| H4 | *There also exists a wealth threshold above which climate mobility is no longer needed or desirable.* | Negative impacts of weather anomalies on migration for households with higher endowments (e.g., asset levels). |

## 3. Data

Our main data source is the recently released *Uniform Panel Dataset* which is derived from the past longitudinal rounds of Nigeria's General Household Survey - Panel (GHS-Panel).[5] The GHS-Panel is a national, longitudinal, multi-topic household survey that has been implemented by the Nigeria National Bureau of Statistics since 2010, in collaboration with the World Bank Living Standards Measurement Study (LSMS). The past GHS-Panel rounds were implemented in 2010-2011, 2012-2013, 2015-2016, and 2018-2019. Each round is composed of two visits, one conducted after crop planting (post-planting visit) and one after the harvest (post-harvest visit). We focus on households who participated in the agricultural questionnaire—i.e., households engaged in agriculture—due to the predominant importance of the agricultural channel in determining climate-migration relationships in low and lower-middle-income countries (Cai et al., 2016; Cattaneo et al., 2019).

We construct our migration outcome variable using survey questions regarding household membership composition across survey visits and waves. The household roster is updated in each wave by asking the main respondent to confirm whether the individuals listed as household members in the previous wave are still part of the household, and then inquiring about new individuals who joined the household since the last survey wave. The same type of roster update is conducted in the post-harvest visit with respect to the post-planting situation. If a previous household member left or is absent from the household, the reason for departure and the current place of residence are asked. This information is important as, clearly, not any absent individual is a migrant: people leave a household for reasons unrelated to migration (e.g., death, split off into another household, etc.). However, it was not possible to fully exploit these data mainly due to substantial inconsistency of responses across waves (e.g., the same episode being attributed to several reasons over the years). In this study, we therefore define an absent individual as a migrant according to the following two criteria: i) he/she was present in the previous wave (or visit) but not in the subsequent one; and ii) his/her current place of residence is either in a different Local Government Area (LGA)[6] or in a different country. The geographic element is introduced to exclude household splits where one or more individuals move to a separate dwelling nearby to live on their own or as part of a newly formed household. The final migration outcome indicator for this study is then aggregated at the household level and consists of a binary variable, which we call 'migrant-sending household', taking value 1 if the household had at least one member who migrated since the previous wave/visit (according to the above definition), and 0 if otherwise.[7] Even though our definition of migrants is general and includes international migrants, the amount of cross-border migrants, as in similar studies, is minimal: 3.6% of the migrant-sending household sample and 0.97% of the full household sample. The empirical results will thus have to be interpreted primarily under the lens of internal migration.[8]

---

[5] The data can be requested [here](.).

---

[6] The LGA is the second administrative level in Nigeria below state (administrative level 1) and above wards (administrative level 3). There are 774 LGAs in Nigeria and their size varies considerably going from a minimum of 4 to a maximum of 11,225 square kilometers, and with an average of 1,175 square kilometers (see here).

[7] The average number of migrant individuals among migrant-sending households is 1.95, and the value ranges from a minimum of 1 (for 51% of observations) to a maximum of 9 individuals per household.

[8] As a robustness test, we also replicated the empirical analysis with an alternative outcome variable including only internal migrants. The results are unchanged and available upon request.

A few remarks are in order regarding our migration variable. First, a general caveat is that it measures whether a household sent one or more migrants in-between waves, but it does not incorporate detailed information regarding the nature of these migration decisions. Therefore, based on the conceptual framework illustrated in Section 2, we are assuming that most of the migration observed in the data is voluntary and economically motivated. To be able to disentangle economically motivated versus involuntary and low-return migration and thoroughly identify the respective transmission channels, we would need more detailed data on migrant individuals e.g., regarding the purpose of their migration, their current location, work status, and financial links with the origin household. Second, despite these limitations, our definition of migration is more precise than other recent studies using household survey data (e.g., Di Falco et al., 2023; Kafle et al., 2020; Mueller et al., 2020) which simply used changes in household size or short-duration absence from the households as proxies for migration. Third, consistent with the conceptual framework outlined in Section 2, it is defined at the household rather than individual level. This is important not only from a conceptual point of view (the household as the decision-making unit, rather than the individual, in line with the NELM perspective), but also for technical and practical reasons due to the rare-event nature of the migration phenomenon. One of the main concerns when using nationally-representative household surveys for studying migration is that the migrant sample size is often (too) small. While this is certainly a limitation if one wants to infer individual-level relationships, it is a less serious problem for household-level statistical analysis. As noted by Carletto et al. (2023), collapsing individual migration data at the household level provides enough of a gain in statistical power and sufficient outcome variation needed for estimation. In our case, we observe just above 3% of migrant *individuals* across the panel (2010-2019) but almost 27% of migrant-sending *households* over the same period.

We then integrate household data with gridded geospatial weather information collected from third-party, publicly-available sources using spatially anonymized household GPS coordinates.[9]

---

[9] Spatial anonymization is standard practice in georeferenced household survey data that are publicly available, to protect the privacy of the respondents. Michler et al. (2022) have recently shown that spatial anonymization methods currently implemented in longitudinal surveys supported by the LSMS do not exert any meaningful impact on estimates of the relationship between weather and agricultural outcomes. In particular, the authors conclude that researchers need not be concerned about potential inaccuracies that may be introduced by integrating spatially-

As a weather indicator, we use the Standardized Precipitation Evapotranspiration Index (SPEI), a multi-scalar drought index developed by Beguería et al. (2014).[10] The SPEI jointly considers precipitation, potential evaporation, and temperature, which provides a distinctive advantage over simple rainfall or temperature indicators as it incorporates the interaction between these variables in determining farmers' agricultural outcomes (Bertoli et al., 2022; Harari & La Ferrara, 2018). It is increasingly used in the literature to capture agriculture-relevant weather shocks and has been found to outperform other indicators in predicting crop yields (Vicente-Serrano et al., 2012). The SPEI is obtained by taking the difference between precipitation ($P$) and potential evapotranspiration ($PET$): $D_i = P_i - PET_i$. Then, $D_i$ is standardized so that the indicator represents the deviation from the normal water balance (a SPEI of 0 indicates a value corresponding to 50% of the cumulative probability of $D$, according to a log-logistic distribution). A negative SPEI value is associated with dry events (lower rainfall) while positive SPEI values capture wet events (higher rainfall).

To capture weather conditions that matter for agricultural outcomes and, in turn, migration decisions, we focus only on growing-season SPEI (GS-SPEI), defined as the average monthly SPEI during the growing season months in the period between two waves. The rationale for this choice is provided by previous literature investigating outcomes and settings where impacts tend to be driven by the agricultural channel (Cai et al., 2016; Cattaneo et al., 2019; Harari & La Ferrara, 2018), which is particularly sensitive to weather variability during the growing season. To identify growing periods, we rely on crop calendars—differentiated by region—provided by the Famine Early Warning System Network (FEWS NET) of the United States Agency for Agricultural Development (USAID).[11] Ideally, one could argue that we should attribute to each migrant-sending household only the average SPEI during the growing season months between the last wave and the month in which the individual migrated. However, this would imply that migration is an

---

anonymized survey datasets with publicly available remote-sensing weather products, since the current spatial resolution of geospatial data is not fine enough for spatial anonymization to result in mismeasurement of weather events that are experienced by sampled households.

[10] The SPEI data are based on gridded data produced by the Climatic Research Unit of the University of East Anglia and can be downloaded here. The level of resolution is 0.5 x 0.5 degrees.

[11] The Nigeria FEWS NET crop calendar is available here.

*ex post* coping strategy, and not an *ex ante* risk management strategy implemented by the household. As noted by Kleemans (2023), the decision to send a household member away can be either in response to a negative growing-season shock or, alternatively, a proactive investment strategy to diversify risk in the face of uncertainty for the upcoming rainy season. For this reason, since we are mainly interested in economic migration, we remain agnostic regarding the timing of the weather shock that matters for the migration move and use as our treatment variable the average SPEI computed over the full set of rainy season months between two waves.[12]

The available evidence suggests that slow-onset events such as droughts retain the largest migration potential compared to fast-onset events such as floods (Cattaneo et al., 2019; Letta et al., 2023). The SPEI measure is able to capture both types of events as large negative or positive deviations from the usual water balance. Usually, when aggregated over a timespan longer than monthly, a drought is characterized by a SPEI value lower than 0 and, symmetrically, positive rainfall shocks are captured by values higher than 0 (Bertoli et al., 2022). Given the known stronger relevance of droughts for mobility decisions, and for the sake of interpretability, we use the reversed value of the GS-SPEI variable, so that an increase of one standard deviation represents a drought event and a decrease of one standard deviation captures heavy rains and floods. Since, in each wave and visit, households are interviewed in different calendar months, we take household-specific averages of GS-SPEI depending on each household's interview date. This means that, for each household, we look at weather conditions occurring in all growing season months between their last and current interview.

In addition to the outcome and treatment variable, we construct several other variables capturing household demographic and socio-economic characteristics to be used as part of our estimation strategy. As we need these variables to be strictly exogenous and pre-treatment (see Section 4), all these characteristics are lagged (with the exception of wave-specific shocks unrelated to climate) and come from information provided by the household interviews in the preceding wave of the panel. Our final (unbalanced) panel dataset includes 5,461 observations of 2,736 agricultural households from waves 2, 3, and 4 of the Nigeria GHS Uniform Panel Data that were also interviewed in the preceding wave. Descriptive statistics for the key variables of interest are reported in Table A1 in the Appendix. Note that the average value of the reversed GS-SPEI is 0.23, suggesting that in our sample drier-than-normal conditions represent the prevalent type of weather anomaly experienced by farming households, and that growing season weather throughout the survey period has been overall less favorable compared to the historical (1900-2020) long-run sample mean over which the SPEI is computed.

---

[12] There are also more practical reasons for this choice: i) the reporting of migration timing is inconsistent in the survey, making the information unreliable; ii) for about half of the migrant-sending households, the number of migrants per wave is higher than one; iii) in the estimation, treated (with migrants) and control (without migrants) households should be assigned the same treatment window, and it is unclear what window should be assigned to households without any migrant if using the weather window before the specific migration move.

## 4. Methods

Our identification strategy is based on the use of causal forests. Causal forests were specifically developed for the estimation of Conditional Average Treatment Effects (CATEs), which constitute non-parametric estimates of treatment effect heterogeneity based on a set of chosen covariates (Athey & Imbens, 2016; Athey et al., 2019; Wager & Athey, 2018). Causal forests are a causal inference adaptation of random forests—a powerful supervised machine learning method aimed at predicting a given outcome on the basis of a set of inputs (Breiman, 2001; Hastie et al., 2009)—to the task of predicting heterogeneity in causal effects.

The causal forest algorithm is an ensemble of causal trees. Each of these trees is defined by data-driven sample splits that generate leaves, which are followed by a prediction of the causal effect over a set of conditioning characteristics *X*. The aim of a causal forest is to split the data so as to maximize treatment effect heterogeneity across leaves. To prevent the risk of overfitting, the algorithm is based on the so-called "honest approach" (Athey & Imbens, 2016), which randomly splits the sample in two equal parts: half of the sample (the *prediction* sample) is employed to define the sample splits (leaves), while the other, the *estimation* sample, is used for estimating the predicted CATE. Such a procedure is repeated as many times as there are trees in the forest (2,000 trees, in our case). Each individual tree explicitly searches for the subgroups where the treatment effects differ most, and the final causal forest prediction is a weighted average over the predictions across trees, which, as Wager and Athey (2018) show, is consistent and asymptotically normal. In order to capture group-level effects, we cluster standard errors of the estimated effects at the

household level.[13]

The difference between this approach and traditional subgroup analysis is that, in the latter, the researcher would define some subgroups of interest and then estimate a CATE within each. The causal forest algorithm removes a key degree of freedom, since the researcher only chooses the set of candidate covariates (i.e., the set of potential drivers of heterogeneity) that can be used to define subgroups, but not the specific subgroups or the critical threshold values used to define them. The algorithm then uses the most predictive among the candidate covariates to partition the sample space and automate the selection of subsamples to favor partitions that have greater variation of treatment effects across subsamples.

Causal forests can be applied on randomized, as well as observational, data. In observational contexts, they have also recently been leveraged to estimate heterogeneous treatment effects in panel data settings (Athey et al., 2023; Britto et al., 2022; Miller, 2020; Zhang & Luo, 2023). Our work belongs to this new strand. The main identifying assumption underlying the causal forest approach is conditional unconfoundedness: to make the assumption credible, we follow Wager and Athey (2018), who recommend a double orthogonalization approach. Specifically, we separately orthogonalize the outcome and treatment variable with respect to potential confounders by running two preliminary and fully non-parametric regression forests in which we include the following in the set of predictors: household fixed effects, non-parametric time trends, household coordinates, LGA dummies, self-reported idiosyncratic shocks (e.g., illness or death of a household member) and man-made shocks unrelated to climate (i.e., robbery, fire, etc.), and lagged demographic characteristics such as age and gender of the household head, household size, and number of children and working-age members in the household.[14] This way, the residualized outcome and treatment variables are filtered from time-invariant household and geographic characteristics, time-varying demographic characteristics, other unrelated shocks, and time trends. Estimation of treatment effects is then carried out using the predicted outcome and treatment

---

[13] For the implementation, we use the *R* package *grf* (Athey et al., 2019). In the estimation, we cross-validate all the tuning parameters.

[14] Household fixed effects are included using the approach for sufficient representation of categorical variables recommended by Johannemann et al. (2019). We employ the *sufrep* R package provided by the authors.

variables obtained from the previous double orthogonalization step.

This approach provides us with multiple layers of robustness to ensure the plausibility of the unconfoundedness assumption. First, weather shocks are generally assumed to be random, and this is why they are often used as instrumental variables (e.g., Kleemans & Magruder, 2018; Miguel et al., 2004).[15] Second, the SPEI is normalized with respect to normal local conditions, thus filtering out heterogeneity due to different local averages. On top of this, the double orthogonalization approach we adopt before treatment effect estimation eliminates any residual source of confounding effects and omitted variable bias. Lastly, all the regression forests are cluster-robust since we explicitly account for household clusters at all estimation steps.

After applying orthogonalization, we proceed with the estimation of CATEs. Specifically, based on the current knowledge and scientific evidence about the key transmission channels of the climate-migration nexus (Cattaneo et al., 2019; Letta et al., 2023), the following vector *X* of observable characteristics enters the CATE estimation model:

- Household asset index[16]
- Lagged GS-SPEI (reversed)
- *In situ* adaptive capacity score[17]
- Annual total consumption per capita (logged)
- Share of adult members with tertiary education

All the conditioning factors are measured at baseline, i.e., in the preceding wave of the panel, so they are rigorously exogenous with respect to the treatment. The asset index (which provides an indication of household's wealth relative to others in the survey) and consumption account for heterogeneity in wealth and liquidity constraints.[18] The difference is that the consumption data capture short- and medium-term deprivation, whereas assets (as also emphasized in the conceptual framework) represent a better indicator of long-term wealth (Aiken et al., 2023). Indeed, the correlation between these two variables is relatively low in our sample (0.37). The lagged values of the reversed GS-SPEI are included to capture lagged responses and cumulative or nonlinear effects due to repeated shock exposure over time.[19] Previous survey literature has made a strong case for investigating the way populations respond to the risk of cumulative shocks (Cattaneo et al., 2019), especially since climate change is the reason for the increased frequency and intensity of extreme weather events (Stott, 2016). The adaptive capacity score is included to capture heterogeneity due to a household's pre-shock ability to adapt *in situ*. This variable is of particular relevance since the dynamics of the relationship between on-farm adaptation and migration remains under-explored (Cattaneo et al., 2019), and is built by leveraging information on farmers' adoption of agricultural inputs and practices that are relevant in adapting to climate risk.[20] The education variable captures the role of human capital of household members in determining the household-level decision of dealing with climate risk by sending out individual migrants, another key mediating channel emphasized in the literature on climate migration (Burzyński et al., 2022) and a main determinant of migration in the NELM framework (Stark & Bloom, 1985).

---

[15] In the climate panel econometrics literature, it is common practice to simply regress the outcome of interest on the set of weather variables plus a full set of unit fixed effects and time dummies or trends (Dell et al., 2014; Hsiang, 2016).

[16] The construction of the asset index follows the approach of Aiken et al. (2023) and is based on principal-component factor analysis. See Table A2 in the Appendix for the list of items included. To ease interpretability, the estimated index has been rescaled to be in the range 0-100.

[17] This score is also built via principal-component factor analysis; see Table A2 in the Appendix for the list of items included. To ease interpretability, the estimated index has been rescaled to be in the range 0-100.

The choice of including only a limited number of conditioning variables for the estimation of CATEs is consistent with the adoption of the double orthogonalization approach discussed above and allows us to focus on those elements that we believe to be true treatment effect modifiers.[21] Furthermore, to obtain valid inference with causal forests, the vector of conditioning variables $X$ should be low-dimensional (Athey et al., 2019). Since our outcome variable is binary, the treatment effect must be interpreted as the change in probability of the outcome occurring that is associated with the treatment (the weather anomaly).

---

[18] Monetary consumption values have been temporally and spatially deflated.

[19] The SPEI variable we built captures growing-season weather anomalies occurring during the entire between-wave interval across different survey rounds. Including a lagged value of this variable means capturing weather conditions over a timespan of several years. For instance, by focusing on wave 4 (2018-2019) and using both current and lagged values of the reversed GS-SPEI, we can capture weather conditions in the period between wave 3 (2015-2016) and wave 4, as well as weather conditions in the period between wave 2 (2012-2013) and wave 3.

[20] For instance, we include the use of pesticides in the score as the agronomic literature indicates that higher temperature and precipitation increase pest pressure, which may in turn lead farmers to use more pesticides (Bareille & Chakir, 2023). This results in a high degree correlation between different adaptive indicators, which is important in our case since we are not able to include all the possible adaptation variables in the construction of the index (e.g., use of drought-tolerant seeds, intercropping practices, etc.) due to missing data.

To enhance the plausibility of our research design, we also carry out a placebo test. In observational studies, placebo tests represent a key device for assessing the credibility of research designs (Egger et al., 2023). To this end, we randomly reshuffle our treatment, i.e., average SPEI values during the growing season months in the period between waves, across the entire sample. Then, we re-run the exact same empirical analysis with this permuted treatment.[22] If we are detecting systematic effect heterogeneity rather than noise, we should expect no treatment effect heterogeneity whatsoever, across any conditioning variable, resulting from this falsification exercise.

## 5. Results

The causal forest estimates indicate a positive but statistically insignificant Average Treatment Effect (ATE) of 5.9% (standard error: 4.1).[23] However, this average effect masks substantial heterogeneity across the spectrum of household responses to weather anomalies, as shown in Figure 2, which reports the distribution of the estimated effects. This heterogeneity concerns not only the magnitude, but even the sign of the impact: weather anomalies increase the probability of sending a migrant for some groups of households but decrease it for others. In contrast to the ATE, the predicted individual effects are statistically significant (at the 5% level) for virtually all households, and range from –38.7% to +61.2% changes in the probability of sending a migrant, corresponding to about –13% and +20.6% for a one standard deviation increase in the treatment (48.3% and 76.6%, respectively, of the unconditional mean of the dependent variable). These results support our argument that focusing on average effects when there is substantial underlying

---

[21] Since double orthogonalization eliminates confounding effects at the onset, there is no need to include variables that could be potential confounders in the estimation stage (Athey & Wager, 2019).

[22] As for the benchmark model, we employ the reversed values of the SPEI variables also in the placebo test.

[23] With causal forests, the ATE is computed using the doubly robust score for average partial effect estimation with continuous treatment recommended by Wager and Athey (2018) and Athey et al. (2019).

heterogeneity can result in misguided policy prescriptions.

For comparison, the ATE estimated for the placebo test is indistinguishable from zero (0.2%) and statistically insignificant (placebo standard error: 1.8). The histogram of Figure A1 in the Appendix shows the corresponding placebo effects, which appear normally distributed and centered around zero, without any relevant heterogeneity. The null ATE and the absence of any effect heterogeneity in this placebo test reinforces the plausibility of the research design and suggests that our estimates are picking up a signal of a real impact of agriculture-relevant weather anomalies on migration decisions.

Where does the substantial heterogeneity depicted in Figure 2 come from? This is the key question addressed in Figure 3, in which, following Britto et al. (2022), we look at the estimated CATEs for each corresponding decile of the selected treatment effect modifiers.[24] For three conditioning variables, which correspond to key elements underlying the dynamics depicted in Figure 1, we see steep gradients in the estimated treatment effects. Specifically, the figure suggests that the largest degree of treatment effect heterogeneity is driven by pre-shock levels of the asset index and local adaptive capacity and repeated shock exposure.

For households within the lowest two deciles of asset values, anomalous weather decreases the probability of sending a migrant by more than 5%, whereas the impact turns positive for households in the upper deciles and is especially pronounced (10%) for better-endowed households in the highest decile of the pre-shock distribution. This positive gradient is consistent with a gradual relaxation of liquidity constraints associated with higher wealth and provides empirical support for the H1 hypothesis about the existence of a climate-induced asset threshold under which households are trapped in immobility (cf. Table 1).[25]

---

[24] For the share of tertiary education, there are only three bins as the distribution of this variable is skewed towards zero (see Table A.1) and it is not possible to bin it into ten deciles.

[25] In an immobility-oriented framework, it is perfectly conceivable that weather shocks might increase migration intentions while, at the same time, impairing actual migration flows. In this regard, a recent cross-country study by Bertoli et al. (2022) finds that weather anomalies do increase migration intentions in Nigeria.

**Figure 2: Treatment effect distribution**

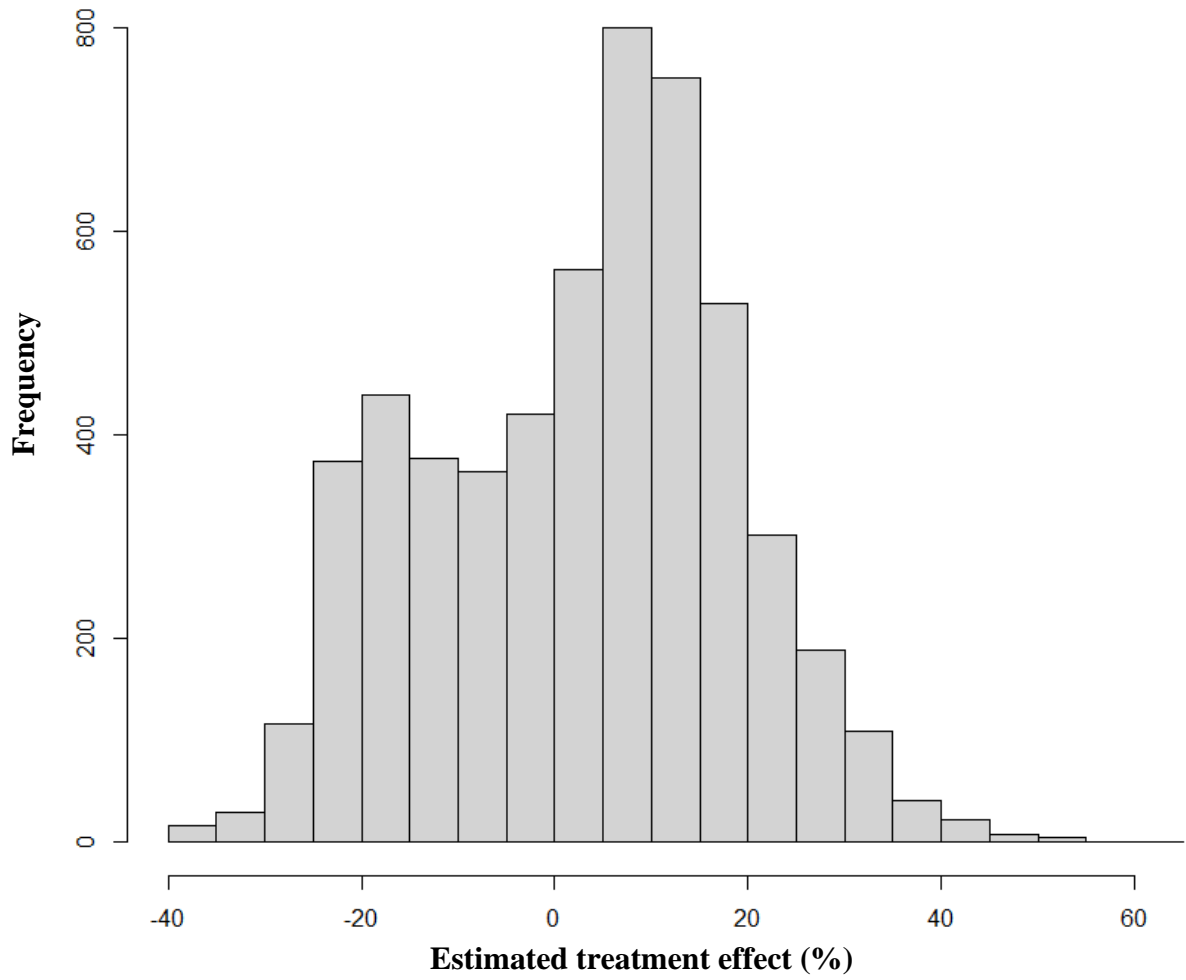

**Figure 3: Conditional Average Treatment Effects (CATEs)**

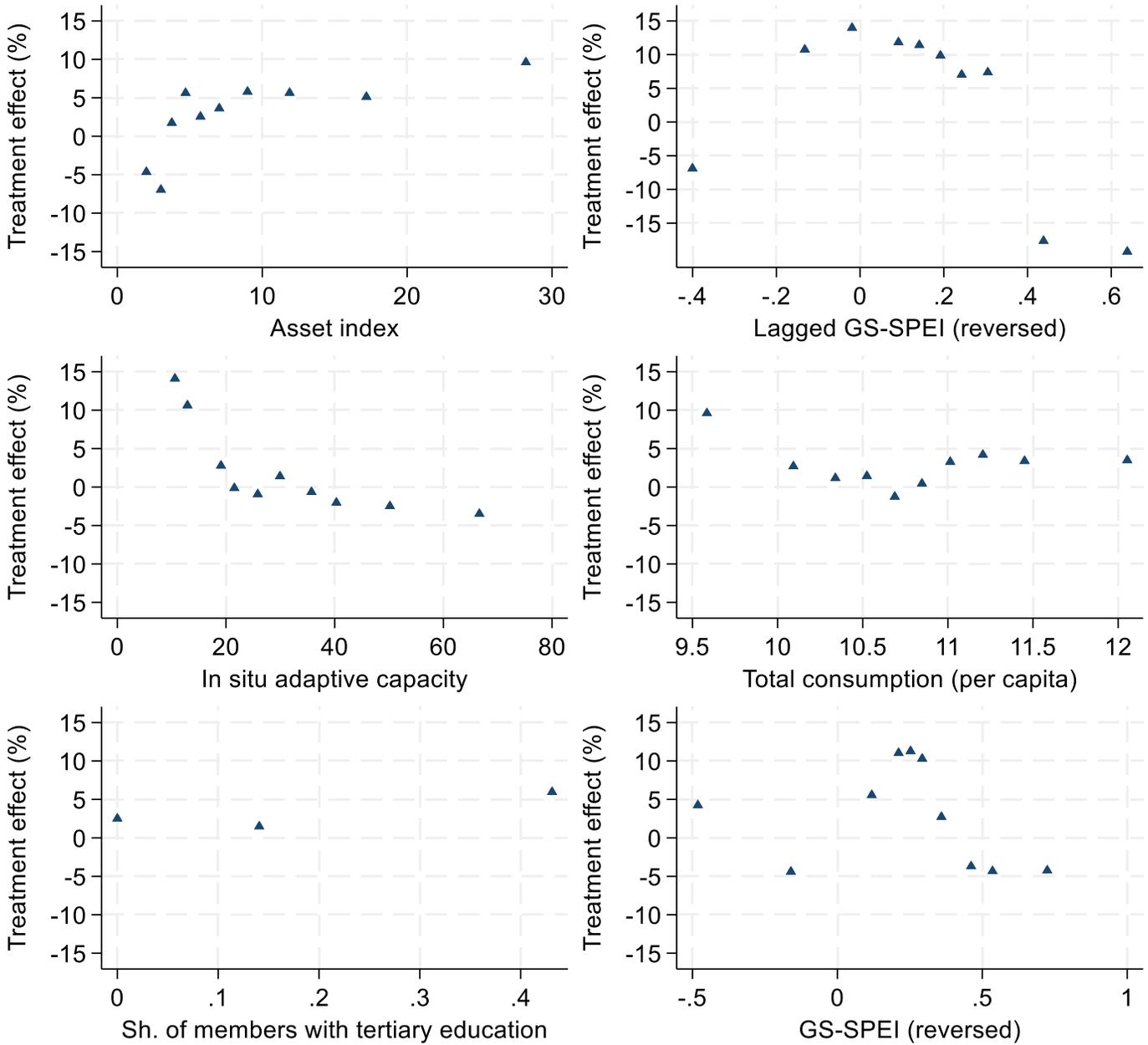

*Note*: the triangles represent average CATE values for each decile of the treatment effect modifier reported on the X-axis. Treatment effect is the change in the probability of a household sending at least one migrant associated with a one standard deviation increase in the reversed GS-SPEI. All the treatment effect modifiers are lagged variables from the preceding wave. Variable 'GS-SPEI (reversed)' shows heterogeneity in treatment effects depending on treatment intensity levels.

Conversely, we do not find empirical support for the H4 hypothesis on the existence of an upper voluntary immobility threshold: we do not observe any reduction in the probability of migration for wealthier households, which would have signaled that sending migrants in anticipation of or in response to shocks is no longer necessary. Therefore, this finding is at odds with the expected result for this hypothesis (negative effects for households with high asset levels). The absence of an inflection point may indicate either that this upper threshold just does not exist for farming households whose income is inextricably linked to biological processes dependent on weather (e.g., agronomic conditions), or that we simply do not observe such level of wealth in our sample.[26]

The relationship of the treatment effects with respect to past weather conditions is concave and nonlinear: previous exposure to drier-than-normal conditions is associated with a negative and sizable (up to −20%) effect of current weather shocks on the probability of being a migrant-sending household. This nonlinear relationship is consistent with hypothesis H2 and the notion advanced in previous literature (Cattaneo et al., 2019) that, more than the occasional exposure to weather shocks, what matters is the recurrent, cumulative exposure to climate stress and severe shocks. Specifically, living in areas recurrently plagued by anomalous weather seems to be conducive to resource-constrained immobility and may give rise to geographic immobility pockets. This result partially contradicts findings from other recent work (Di Falco et al., 2023), which instead found that cumulative drought exposure leads to a significant increase in the probability of migration.

Other than assets and previous shocks, there is a steep negative gradient with respect to *in situ* adaptive capacity: households with a low adaptive capacity score tend to send more migrants due to rainfall scarcity (with an effect close to +15%), whereas the impact turns close to zero for households with higher adaptive capacity, until becoming negative for the most adaptive households. This suggests that migration and local adaptation are substitutes rather than complements: households with more irrigated plots, making more use of modern agricultural inputs, and with off-farm enterprises are less likely to send migrants due to dry spells. This is in line with hypothesis H3 about substitutability between *in situ* adaptation and migration as alternative risk mitigation strategies for farming households.

---

[26] In our sample, even better-endowed farming households are relatively poor: a back-of-the-envelope calculation reveals that, after accounting for Purchasing Power Parity (2017) and converting local currency to US dollars, the average daily per capita consumption for households in the highest asset index decile is $4.22. While this figure is above the international line of $2.15 (based on 2017 PPPs) for extreme poverty currently used by the World Bank, and also slightly above the $3.65 poverty line set for lower-middle-income countries such as Nigeria, it is below the $6.85 per day poverty line of upper-middle-income-countries. For more information on poverty lines, see here.

In contrast, the role of consumption is not as strong as the one played by previous variables. The relationship with treatment effects distribution is almost flat, with a pertinent exception: households with the lowest absolute level of consumption per capita are more likely to send climate migrants (this effect is close to 10%). Remember that, compared to the asset index, the consumption data capture short- and medium-term deprivation, whereas the asset index is a better indicator of long-term wealth (Aiken et al., 2023). While one might be tempted to interpret this particular result as a hint that the poorest households, unable to smooth consumption, tend to engage in risk-coping migration, this is not the case: upon further investigation, we found that these households simply experienced substantially milder weather conditions compared to the rest of households. The GS-SPEI value for this lowest consumption decile is 0.11, against a sample mean of 0.23 (cf. Table A1). Therefore, this positive effect should be interpreted as the casual result of minor shock exposure rather than as a signal of an underlying heterogeneity related to the conditioning variable. Concerning human capital, we do not observe relevant heterogeneity with respect to the share of members with higher education. Finally, the last panel of Figure 3 shows variation depending on the intensity of the treatment, the reversed GS-SPEI. As this is a continuous treatment, it is interesting to inspect *ex post* the distribution of treatment effects with respect to the treatment 'dose', i.e., the intensity and sign of the experienced weather conditions. The relationship appears strongly nonlinear, and drier conditions (upper deciles of the reversed GS-SPEI) are associated with larger negative effects.

Now note the difference between these CATEs and the estimated placebo CATEs which are reported in the Appendix, Figure A2. The figure illustrates the distribution of the placebo treatment effects for the same set of variables. All the implied relationships are flat and around zero, pointing to absence of any type of heterogeneity in the placebo effects. This reinforces our point that what we find in Figure 3 is systematic heterogeneity and not noise.

Given the predominant role played by the three main heterogeneity drivers—lagged asset levels,

adaptive capacity, and previous weather shocks—we also investigate how these key conditioning variables interact with each other in determining patterns of heterogeneity. Figure 4 reports the three 'heatmaps' of the mean predicted treatment effects over pairs of these characteristics. From the figure we can derive three main additional insights. First, households with the highest positive effects are those characterized by a combination of high pre-shock levels of assets and low local adaptive capacity (Panel A). Second, while being wealthier or less adaptive *in situ* is generally associated with large positive effects, the trapping effect of cumulative shock exposure dominates over both mediators (Panels B and C). The implication is that climate-induced immobility also represents a risk for better-endowed but highly exposed households. Third, Panel D, which overlaps treatment intensity with previous weather conditions, reveals that, regardless of current shock intensity levels, having experienced severe droughts in the previous wave turns the treatment effects negative, suggesting that past shocks matter even more than contemporaneous weather conditions in determining (im)mobility outcomes, which is consistent with the conceptual framework and the notion of geographic poverty traps (Kraay & McKenzie, 2014; Jalan & Ravallion, 2002).

Overall, our empirical test confirms three of the four hypotheses set out in Section 2. Whereas no voluntary immobility threshold has been detected, the test provides evidence that migration and local adaptation are substitute rather than complementary mechanisms, and that households characterized by limited assets and cumulative shock exposure are those at a higher risk of immobility, as they can neither adapt *in situ* nor migrate.

**Figure 4: Treatment effect heatmaps**

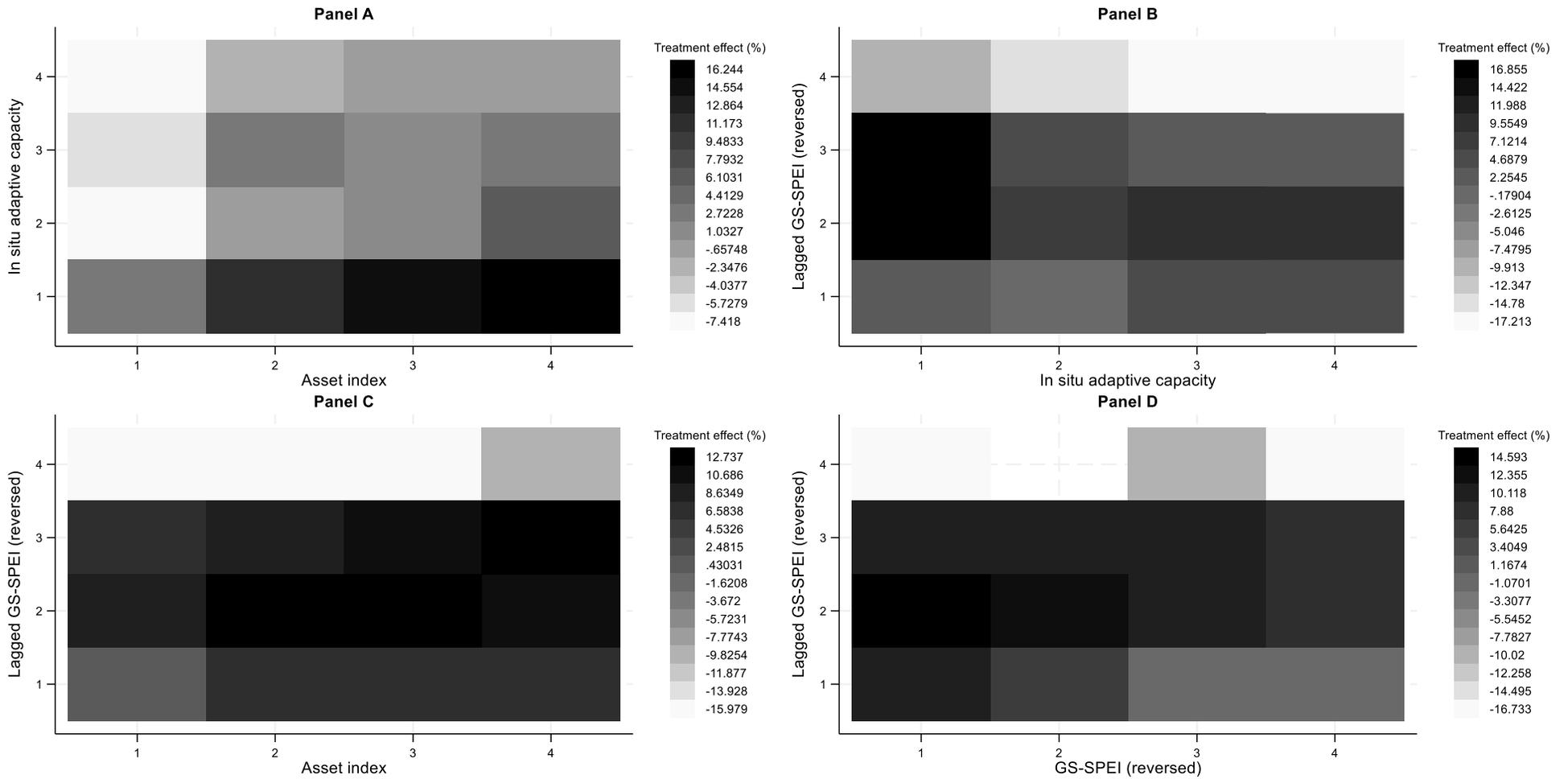

*Note*: the values within each cell represent the mean predicted treatment effects for observations belonging to that cell. Each cell is defined by the corresponding combination of conditioning variable quartiles.

## 6. Conclusions

Despite the growing policy relevance of climate migration and the acceleration of climate change impacts, our understanding of the causal relationship between climate shocks and human mobility outcomes remains limited. While this is in part due to the existence of severe data gaps, there is much that can be done to inform policymaking, even with the current data. In this paper, we carried out an empirical test which shows that coupling rich household survey data and theoretical insights with recently developed data-driven techniques is a promising route that can shed new light on the key drivers underlying heterogeneous mobility responses in the face of climate risk and help target groups most at risk of resource-constrained immobility.

We find that pre-shock wealth, *in situ* adaptive capacity, and repeated exposure are the most important drivers able to determine not only the magnitude but even the sign of the effects of weather anomalies on the (im)mobility outcomes of Nigerian farming households. While local adaptive capacity acts as a substitute for migration, the importance of climate-induced immobility is consistent with recent global evidence that many people will be trapped in impoverished and troubled regions and that climate-induced poverty is likely to be a real threat (Burzyński et al., 2022). The inability to migrate in reaction to climate change is an important and neglected policy concern (Castells-Quintana et al., 2018) and stresses the need for policies that facilitate both local adaptation as well as more internal migration to lift vulnerable people out of poverty traps (Kraay & McKenzie, 2014). If there is an inflection point at which migration starts to decrease as households get substantially wealthier and, in turn, become able to adapt *in situ*—i.e., a kind of 'climate migration Kuznets curve'—we do not observe it, most likely because even relatively richer Nigerian farming households are still poor.

In terms of policy prescriptions, our findings suggest that, in this and similar contexts, any one-size-fits-all approach for the management of climate migration is doomed to fail. The different groups we identify call for targeted measures and group-specific programs and adaptation policies able to tackle the complex challenges stemming from climate-mobility interactions at multiple levels. We propose a methodology for group decomposition and threshold identification which is based on public, nationally-representative microdata available to policymakers and that can be easily tailored to different contexts. As such, this type of empirical approach for detecting treatment effect heterogeneity can be effectively leveraged to fine-tune optimal policy targeting.

Several caveats are in order regarding the limitations of this work. The most important one is the use of a broad definition of migration for our migration outcome variable which is driven by limitations in the existing data, as discussed in Section 3. To overcome this constraint, we stress the need for longitudinal household surveys to collect more comprehensive migration data. This includes details about the nature and duration of migration episodes, as well as the characteristics, history, and purposes of migrant individuals. Achieving this goal requires equipping household surveys with improved migrant tracking tools and protocols. Ideally, efforts to enhance the design of longitudinal surveys should be also harmonized and standardized across countries to ensure comparability of migration data and enable researchers to study potential differences in country-specific relationships. At the same time, for half of the migrant-sending households, we observe more than one household member migrating across waves, so even if it were possible to distinguish different types of migration, accounting for intra-household heterogeneity in the estimation approach would not be straightforward. A complementary strategy to investigate resource-constrained immobility would be to look at the dynamics of migration intentions and aspirations rather than actual migration, but the available information on these variables is limited to cross-sectional surveys (e.g., Gallup World Poll) which come with their related econometric challenges. Regarding internal versus international migration, while our results apply predominantly to internal migration, we suspect that external validity with respect to international migration is high, as the key finding regarding the importance of endowments and liquidity constraints should hold even truer when it comes to costlier cross-border movements.

Secondly, it must be emphasized that conducting this type of research is not possible without leveraging panel surveys that are multi-topic and interoperable in nature. Unfortunately, such surveys remain uncommon in many developing contexts, which hinders the possibility of studying climate migration dynamics in those regions. Relatedly, even in countries where panel data are available, these are often of limited length preventing the analysis of longer-term dynamics and impacts. This is also the case of this paper, where we can only provide short-run evidence in support of our conceptual framework. Having longer panels would allow to go a step further and establish whether the short-run relationships detected ultimately lead to permanent immobility traps. In this respect, our recommendation is that existing longitudinal survey systems, including those supported by the World Bank LSMS, should be sustained to span longer-term panels, and similar systems should be set up and maintained in countries and regions that lack them and that also face increasing scope, intensity, and frequency of climatic shocks.

Lastly, climate migration is and will always be inherently context-dependent: the evidence we have provided for Nigeria has limited external validity. Replicating this kind of analysis on more countries—ideally using harmonized cross-country migration data—appears essential for broadening the scope of our knowledge on this multifaceted phenomenon.

Beine, M., Noy, I., & Parsons, C. (2021). Climate change, migration and voice. *Climatic Change*, *167*(1-2), 8.

Benveniste, H., Oppenheimer, M., & Fleurbaey, M. (2022). Climate change increases resource-constrained international immobility. *Nature Climate Change*, *12*(7), 634-641.

Bertoli, S., Docquier, F., Rapoport, H., & Ruyssen, I. (2022). Weather shocks and migration intentions in Western Africa: Insights from a multilevel analysis. *Journal of Economic Geography*, *22*(2), 289-323.

Boas, I., Farbotko, C., Adams, H., Sterly, H., Bush, S., Van der Geest, K., ... & Hulme, M. (2019). Climate migration myths. *Nature Climate Change*, *9*(12), 901-903.

Breiman, L. (2001). Random forests. *Machine learning*, 45, 5-32.

Britto, D. G., Pinotti, P., & Sampaio, B. (2022). The effect of job loss and unemployment insurance on crime in Brazil. *Econometrica*, *90*(4), 1393-1423.

Burzyński, M., Deuster, C., Docquier, F., & De Melo, J. (2022). Climate change, inequality, and human migration. *Journal of the European Economic Association*, *20*(3), 1145-1197.

Cai, R., Feng, S., Oppenheimer, M., & Pytlikova, M. (2016). Climate variability and international migration: The importance of the agricultural linkage. *Journal of Environmental Economics and Management*, *79*, 135-151.

Cai, S. (2020). Migration under liquidity constraints: Evidence from randomized credit access in China. *Journal of Development Economics*, *142*, 102247.

Carletto, C., & Letta, M., & Montalbano, P., & Paolantonio, A., & Zezza, A. (2023). Too rare to dare? Leveraging household surveys to boost research on climate migration. *World Bank Policy Research Working Paper*, n. 10613.

Carter, M. R., & Barrett, C. B. (2006). The economics of poverty traps and persistent poverty: An asset-based approach. *The Journal of Development Studies*, *42*(2), 178-199.

# Appendix

## Table A1: Descriptive statistics

| Variable name | Mean | SD | Min | Max |
| --- | --- | --- | --- | --- |
| Migrant-sending household | 0.269 | 0.443 | 0 | 1 |
| Growing-season SPEI (reversed) | 0.229 | 0.337 | -0.800 | 0.999 |
| Asset index | 13.308 | 12.553 | 0 | 100 |
| *In situ* adaptive capacity | 32.230 | 18.305 | 0 | 100 |
| Annual total consumption per capita (log) | 10.774 | 0.692 | 8.537 | 15.142 |
| Share of adult members with tertiary education | 0.046 | 0.145 | 0 | 1 |
| Age of the household head | 52.982 | 14.406 | 17 | 108 |
| Gender of the household head (Yes = 1) | 0.135 | 0.341 | 0 | 1 |
| Household size | 7.607 | 3.572 | 1 | 33 |
| Number of children | 2.919 | 2.397 | 0 | 18 |
| Number of working-age members | 3.435 | 2.013 | 0 | 23 |
| Idiosyncratic shock (Yes =1) | 0.160 | 0.366 | 0 | 1 |
| Man-made shock (Yes = 1) | 0.086 | 0.281 | 0 | 1 |

**Table A2: List of variables employed for the construction of the indexes**

| Asset Index | *In situ* Adaptive Capacity |
|---|---|
| Furniture (3/4 Piece Sofa Set) | Household owns a non-farm enterprise (Yes=1) |
| Furniture (Chairs) | Share of irrigated plots |
| Furniture (Tables) | Share of plots on which herbicides were used |
| Mattress | Share of plots on which inorganic fertilizers were used |
| Bed | Share of plots on which organic fertilizers were used |
| Mat | Share of plots on which pesticides were used |
| Sewing machine | Total number of plots owned by the household |
| Gas cooker | |
| Stove (Electric) | |
| Stove Gas (Table) | |
| Stove (Kerosene) | |
| Fridge | |
| Freezer | |
| Air conditioner | |
| Washing machine | |
| Electric clothes dryer | |
| Bicycle | |
| Motorbike | |
| Cars and other vehicles | |
| Generator | |
| Fan | |
| Radio | |
| Cassette recorder | |
| Hi-Fi (Sound system) | |
| Microwave | |

| Iron | |
|---|---|
| TV set | |
| Computer | |
| DVD Player | |
| Satellite dish | |
| Musical instrument | |

*Note*: all items used for the asset index are dummy variables which take value 1 if the household owns at least one of the corresponding asset and 0 otherwise. Variables used for the adaptation score are all shares with the exception of 'non-farm enterprise', which is a dummy variable taking value 1 if the household owns a non-farm enterprise a 0 otherwise.

**Figure A1: Placebo treatment effect distribution**

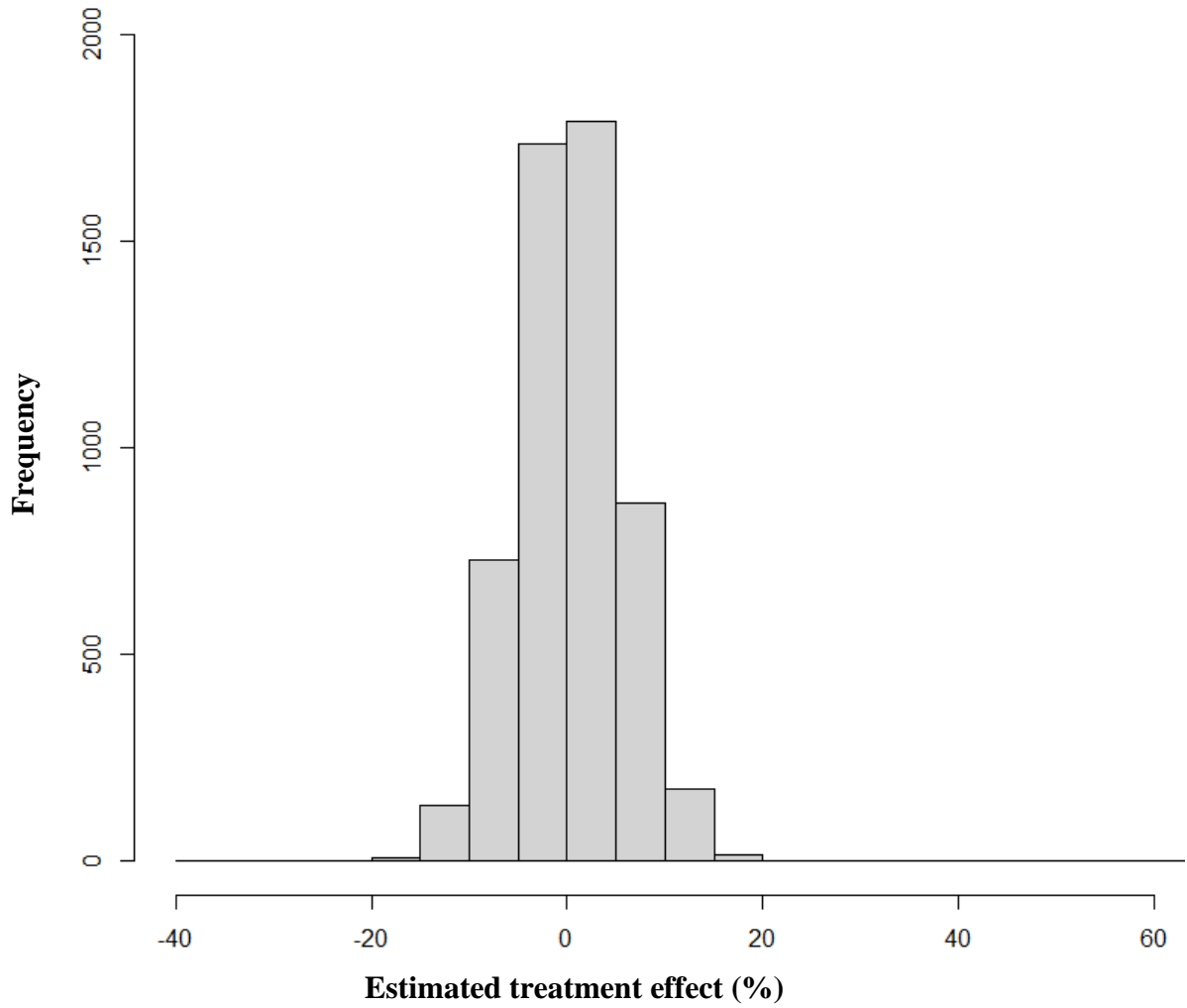

## Figure A2: Placebo CATEs

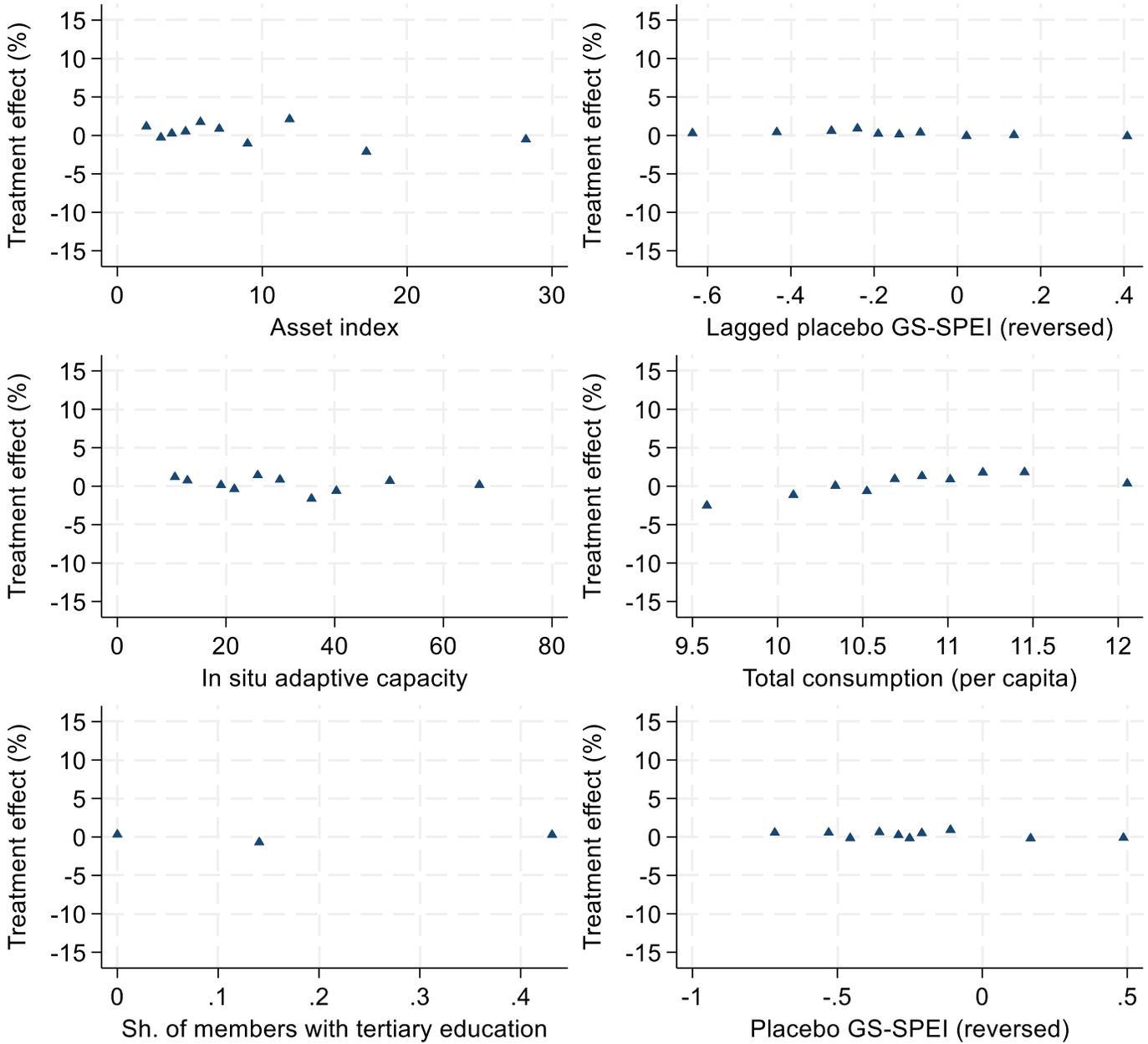

*Note*: the triangles represent average CATE values for each decile of the placebo treatment effect modifier reported on the X-axis. Treatment effect is the change in the probability of a household sending at least one migrant associated with a one standard deviation increase in the reversed placebo GS-SPEI. All the treatment effect modifiers are lagged variables from the preceding wave. Variable 'Placebo GS-SPEI (reversed)' shows heterogeneity in treatment effects depending on placebo treatment intensity levels.